\documentclass{emulateapj}

\usepackage{natbib}  
\usepackage{times}
\usepackage{color}
\usepackage{graphicx} 
\definecolor{RoyalBlue}{rgb}{0.25,0.41,0.88}

\shorttitle{Shadows in the HD~142527 disk} \shortauthors{Marino, Perez
  \& Casassus}

\begin{document}




\title{Shadows cast by a warp in the  HD~142527 protoplanetary disk}


\author{S. Marino\altaffilmark{1,2}, S. Perez\altaffilmark{1,2} and
S. Casassus\altaffilmark{1,2}} \affil{$^1$ Departamento de
  Astronom\'ia, Universidad de Chile, Casilla 36-D, Santiago, Chile}
\affil{$^2$ Millenium Nucleus ``Protoplanetary Disks in ALMA Early
  Science,'' Universidad de Chile, Casilla 36-D, Santiago, Chile}
\email{smarino@das.uchile.cl}

\begin{abstract}

Detailed observations of gaps in protoplanetary disks have revealed
structures that drive current research on circumstellar disks.  One
such feature is the two intensity nulls seen along the outer disk of
the HD~142527 system, which are particularly well traced in polarized
differential imaging. Here we propose that these are shadows cast by
the inner disk.  The inner and outer disk are thick, in terms of the
unit-opacity surface in $H$-band, so that the shape and orientation of
the shadows inform on the three-dimensional structure of the
system. Radiative transfer predictions on a parametric disk model
allow us to conclude that the relative inclination between the inner
and outer disks is 70$\pm$5~deg. This finding taps the potential of
high-contrast imaging of circumstellar disks, and bears consequences
on the gas dynamics of gapped disks, as well as on the physical
conditions in the shadowed regions.

\end{abstract}


\section{Introduction}

The study of planet formation is rapidly being revolutionized by
observational breakthroughs, such as the direct images of
protoplanetary gaps in near-infrared scattered light. A particularly
interesting target is the Herbig Ae/Be star HD~142527, at a distance
of $\sim$140~pc. The disk surrounding HD~142527 boasts the largest
inner cavity known, with a 1\arcsec~ radius and seen at a close to
face-on orientation, with an inclination of about 24~deg
\citep{Fukagawa2006, Fujiwara2006}. Improvements in $K_{\rm s}$-band
coronagraphy revealed fine structure in the outer disk, including two
intriguing intensity nulls \citep{Casassus2012ApJ...754L..31C} that
break the continuity of the outer disk. In this Letter we propose that
these are shadows cast by a warped inner disk.

Recent polarized differential imaging (PDI) of HD~142527 at $H$- and
$K_{\rm s}$-bands revealed a striking view of its dust-depleted gap,
outer disk and spiral structures \citep{Canovas2013A&A...556A.123C,
  Avenhaus2014ApJ...781...87A, Rodigas_2014ApJ...791L..37R}. These
$H$-band polarized intensity images outline the outer-disk nulls with
the best available detail thus far. One null is found $\sim$11.5h
(North of the star), and the other is found at 7h (South). The
Northern null is particularly puzzling as it coincides with the
location of the dust density peak, whose spatial distribution is shaped
into a large crescent \citep{2008Ap&SS.313..101O,Casassus2013Natur,
  2013PASJ...65L..14F}. Thus the Northern null cannot be interpreted
as a lack of material.

As witnessed by a bright thermal IR central point source, well in
excess of the photospheric emission, the large dust-depleted cavity of
the HD~142527 disk is in fact a gap \citep{Fujiwara2006,
  Verhoeff2011A&A...528A..91V, Rameau2012A&A...546A..24R}.  While
knowledge of the outer edge of the gap is limited by theory rather
than observational detail, the inner regions of the gap are poorly
resolved.  A $\sim 10^{-7}$~$M_\odot$~yr$^{-1}$ stellar accretion
rate \citep{2006A&A...459..837G} would quickly deplete the inner disk,
whose dust mass is estimated at $\sim 10^{-9}$~$M_\odot$
\citep{Verhoeff2011A&A...528A..91V}. This inner disk can be
thought of as a transient feature of stellar accretion: it is the
convergence point of matter being accreted from the mass reservoir in
the outer disk \citep{Casassus2013Natur, Casassus2013rocks,
  Casassus2015twist}. Long-baseline optical interferometry (VLTI) has
resolved this inner disk, and found it to be highly crystalline
\citep{2004Natur.432..479V} ---but its orientation and extent remain
elusive.

Motivated by our research on the intra-cavity gas kinematics
\citep{Casassus2013rocks,Casassus2015twist} we considered the
possibility of a continuous warp linking different orientations between
the outer and inner disks. The underlying parametric disk model is
documented in Section~\ref{sec:model}, along with the radiative transfer
setup we implemented to calculate the emergent specific intensity
fields. Our results are presented in Sec.~\ref{sec:results}. While the
inner disk intercepts stellar light, shadows are cast on the outer
disk such that their shape and orientation are sensitive measures of
the inner disk orientation. We briefly consider the consequences of
this finding in Sec.~\ref{sec:discussion}.

\section{Parametric modeling} \label{sec:model}

\subsection{Underlying physical structure}

We constructed a synthetic model of HD~142527 inspired by
multiwavelength observations and gas kinematics, and detailed enough
to provide a comparison point with the near-IR scattered light images
as well as the radio continuum emission and gas kinematics. As
explained in a companion paper \citep{Casassus2015trap}, this
synthetic model matches the observed spectral energy distribution
(SED). Here we document the aspects of the model that are relevant to
this report on the $H$-band images. 

\begin{figure}
\begin{center}
  \includegraphics[width=\columnwidth]{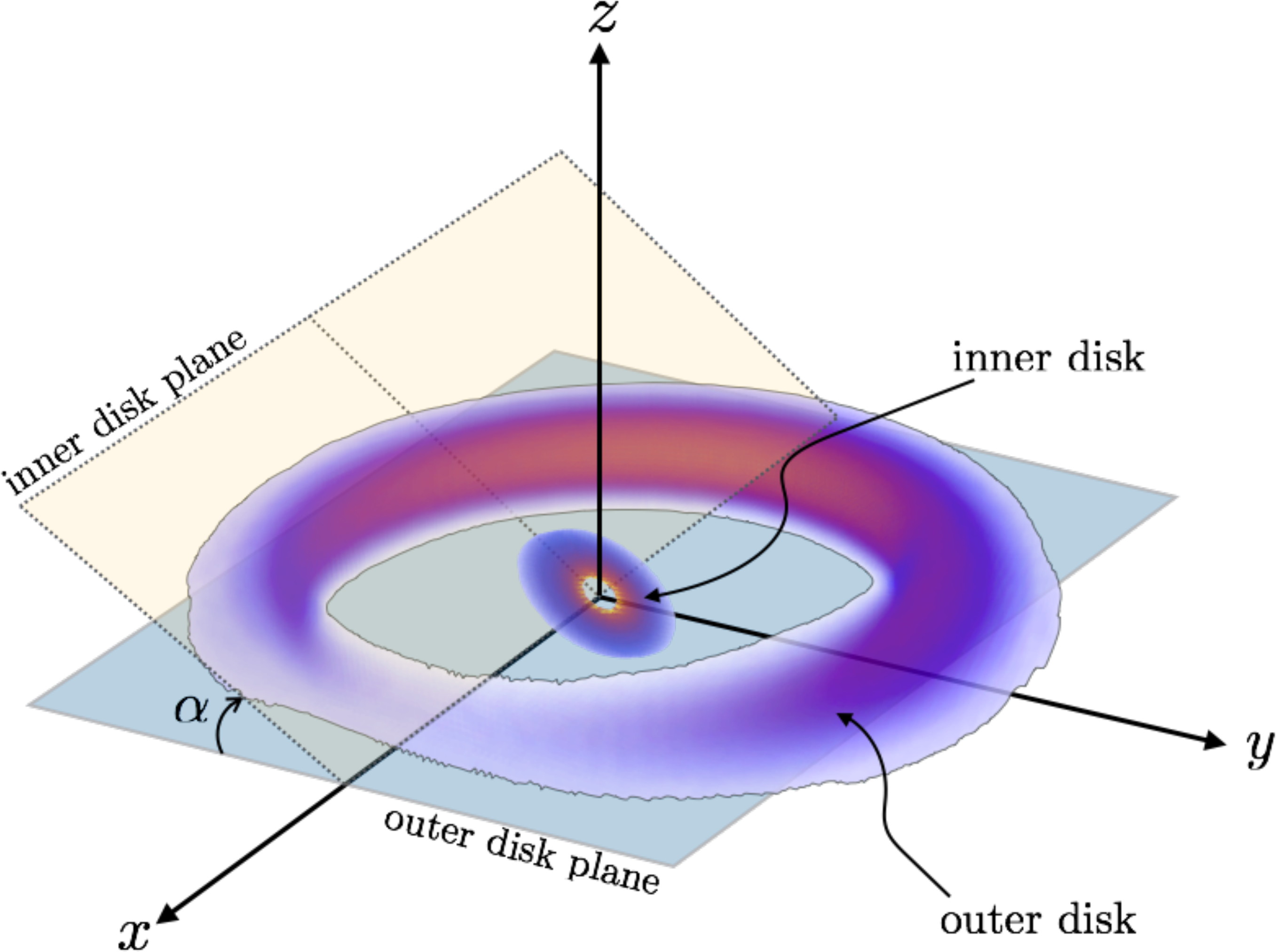}
\end{center}
\caption{ \label{fig:sketch} Schematic view with arbitrary orientation
  of the parametric model presented in Sec.~\ref{sec:model}. The
  central star is placed at the origin. The outer disk lies in the
  $x$-$y$ plane. The angle $\alpha$ is the relative inclination
  between the midplane of the outer disk and the plane of the inner
  disk.  The dust mass density distribution of the inner disk and
  outer disk sections are rendered in false color.  The gap is shown
  devoid of material for simplicity. The inner disk is scaled up in
  size and density for better visualization. }
\end{figure}

The model consists of a star at the center of a dusty disk separated
in 3 zones with different density structures, dust grain size
distributions and compositions: an inner disk, a gap and an outer
disk. A schematic render of the model is provided in
Figure~\ref{fig:sketch}. We model the star using a Kurucz template
spectrum \citep{2003IAUS..210P.A20C} at a temperature of 7750~K and
with a radius of 3.3~$R_{\odot}$, to be consistent with the extinction
of $A_{R}=1.3\pm 0.3$ reported in \citet{Close2014}.

The inner disk starts at 0.3~au and cuts off at 10.0~au. It is
described by a surface density proportional to $r^{-1}$ and a scale
height of 1.0 au at 10 au with a flaring exponent of 1.1. It is
composed of amorphous carbon and silicate grains with sizes between
0.1 to 2.5 $\mu$m, with a total dust mass of 5.0$\times10^{-9}$
M$_{\odot}$. We introduce an inclination of $\alpha$=70$^{\circ}$ to
this region with respect to the outer disk midplane.

The gap spans 120~au from 10.0~au. It is described by a surface
density proportional to $r^{-1}$ and a scale height of 18 au at 100 au
with a flaring exponent of 1.6. It is composed of amorphous carbon and
silicate grains with sizes between 1.0 to 10.0 $\mu$m. The total dust
mass of this section is 1.0$\times10^{-8}$ M$_{\odot}$. The disk in
this section connects the inner and outer regions varying the
inclination linearly from 70$^{\circ}$ to 0$^{\circ}$ between 10 and
15~au, where it matches the outer disk orientation. A larger warp
would have been obvious in the $^{12}$CO kinematics inside the gap
\citep{Perez_S_2014arXiv1410.8168P}, with a concomitantly larger
region where the inclination crosses the plane of the sky.

Finally, the outer disk extends over 115~au to 300~au with a rounded
disk wall between 115 to 140 au. It is composed of 3.0$\times10^{-6}$
M$_{\odot}$ of amorphous carbon grains with sizes ranging from 1 to 10
$\mu$m and 1.0 $\times10^{-2}$ M$_{\odot}$ of silicate grains with
sizes ranging from 100 $\mu$m to 5 cm.  The dust masses inferred in
our models are biased by the lack of grain porosity. The resulting
dust masses are also directly affected by uncertainties on the
internal densities.


We assume that the small grains that account for the bulk of the
near-IR opacity approximately follow the gas background, and we define
an axisymmetric gas distribution with a rounded disk wall 
\citep{2013A&A...557A..68M, 2006ApJ...641..526L} described by
the following surface density: 
\begin{eqnarray}
\bar{\Sigma}_{g}(r<r_c)&=&\Sigma_{c} \left(\frac{r}{r_{c}} \right)^{-\gamma}\exp\left[ - \left(\frac{1-r/r_c}{w} \right)^3 \right], \\
\bar{\Sigma}_{g}(r\ge r_c)&=&\Sigma_{c} \left(\frac{r}{r_{c}} \right)^{-\gamma},
\end{eqnarray}
where $\gamma=6$, $w=0.1$ and $r_c=148.0$ au.  Then we modulate this
distribution to create a maximum gas pressure in azimuth, which is
described by the following equations:
\begin{eqnarray}
\Sigma_{g}(r,\phi)&=&\bar{\Sigma}_{g}(r)\left[1+A(r)\sin\left(\phi+\frac{\pi}{2}\right) \right], \\
A(r)&=& \frac{c-1}{c+1}\exp\left[-\frac{(r-R_{s})^2}{2H^2}\right],
\end{eqnarray}
with $R_s=148$ au and where the azimuthal contrast in surface density
is set to $c=10.0$. The volume density follows with a standard
vertical Gaussian distribution:
\begin{equation}
\rho_{g}(r,z,\phi)=\frac{\Sigma_g(r,\phi)}{\sqrt{2\pi}H}\exp\left[-\frac{z^2}{2H^2} \right],
\end{equation}
with $H(r)=20.0 \left(r/ (130 \ \mathrm{au})\right)^{1.17}$. The exact
value of this flaring exponent is not well constrained.

This parametric model also includes the effect of dust trapping,
following the procedure described in
\citet{Birnstiel2013A&A...550L...8B} and
\citet{Pinilla2012A&A...538A.114P}.  However, the bulk of the opacity
in $H$-band is driven by particles well below the sizes required for
efficient aerodynamic coupling, and so the effects of dust trapping in
the outer disk are not relevant to this report. The runs detailed in
Sec.~\ref{sec:results} confirm that the outer disk is optically thick
at H-band and that the scattered light does not trace the crescent
shape seen in the submm.


\subsubsection{Emergent intensities}

We use {\sc radmc3d}\footnote{http://www.ita.uni-heidelberg.de/
  dullemond/software/radmc-3d/} for radiative transfer computations
\citep[version 0.38,][]{RADMC3D0.38}. Scattering and polarization for
last scattering are treated with scattering matrices for our different
dust species, each one with a power law distribution in grain sizes
with exponent $-3.5$.  To compute the full dust opacity and scattering
matrices we made use of complementary codes in RADMC3D and a code from
\cite{BohreHuffman1983} for ``Mie solutions'' to scattering by
homogeneous spheres. We used the optical constant tables for amorphous
carbon grains from \cite{Li_Greenberg_1997A&A...323..566L}, and for
silicate grains we used \cite{Henning_Mutschke_1997A&A...327..743H}.





We implemented our model in RADMC3D using spherical coordinates, with
regular spacing for the azimuthal angle, and logarithmic spacing in
radius and colatitude (polar coordinate).  Thus, the grid is naturally
refined near the inner disk and the midplane.  The radial grid is
additionally refined as it approaches the inner wall of the outer disk
(near 140 au) to ensure a gradual transition from the optically thin
gap to the optically thick outer disk.  We used $10^{6}$ cells in
total, half of them covering the inner disk and gap, and the rest
sampling the outer disk.  The number of points in the radial,
azimuthal and polar grid meshes is 100 each.

As proposed by \citet{Fujiwara2006}, the Eastern side is probably the
far side since it is broader and brighter in the thermal IR. This
orientation also implies that the observed IR spiral pattern is
consistently trailing \citep{Fukagawa2006,
  Casassus2012ApJ...754L..31C, Canovas2013A&A...556A.123C,
  Avenhaus2014ApJ...781...87A}, even in their molecular-line
extensions into the outer disk \citep{Christiaens2014}. Hence we
calculated the synthetic $H$-band images by inclining the system at
24~deg with respect to the plane of the sky\footnote{the outer disk
  defines the plane of the system}, along a position angle (PA) of
--20~deg.



\begin{figure*}[t]
\begin{center}
\includegraphics[width=\textwidth,height=!]{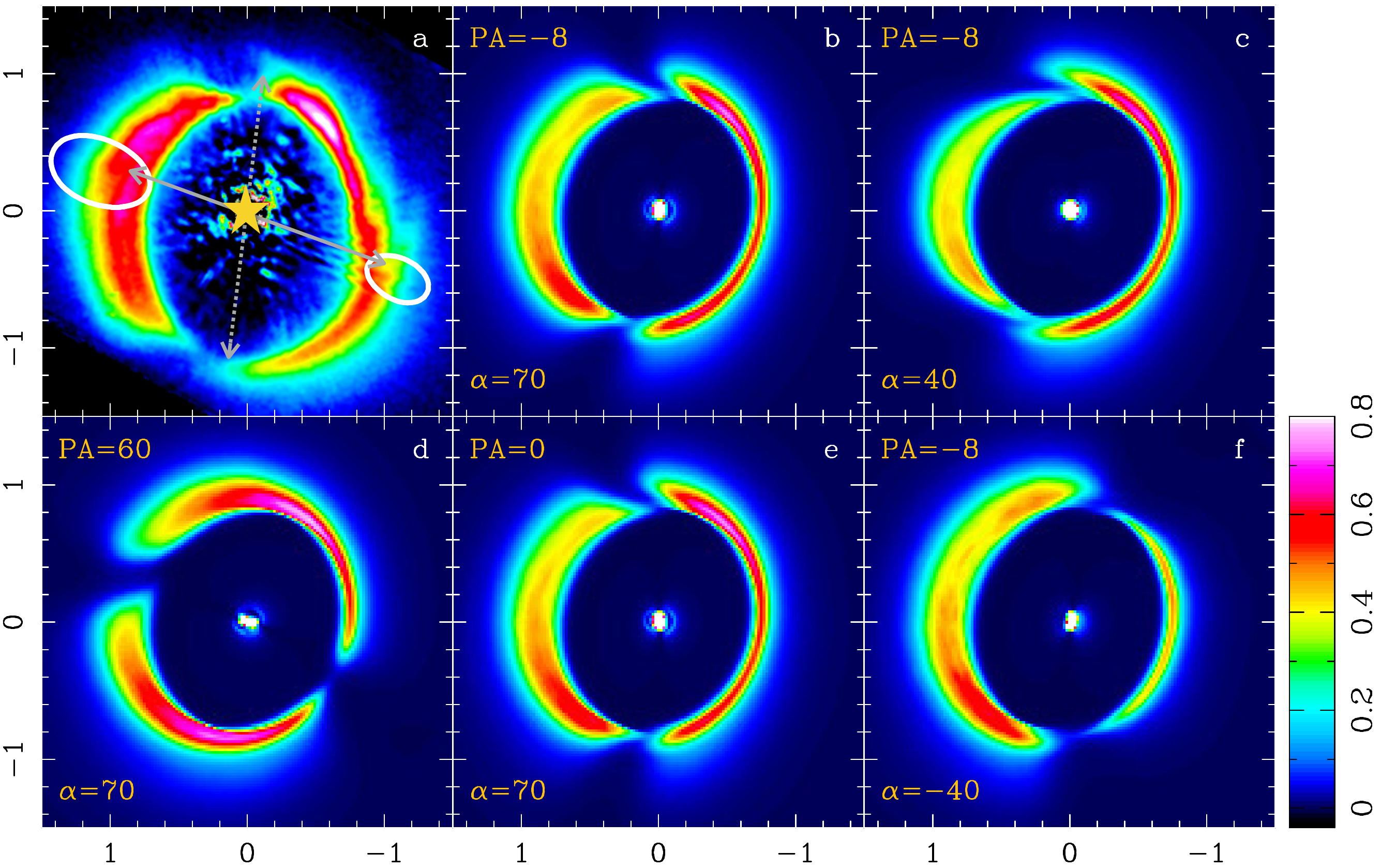}
\end{center}
\caption{ \label{fig:warporient} Impact of the inner disk orientation
  on the H-band light scattered off the outer disk. {\bf a:} NACO-PDI
  H-band image from \citet{Avenhaus2014ApJ...781...87A}, compared with
  the C$^{18}$O(2-1) emission at systemic velocity from
  \citet{Perez_S_2014arXiv1410.8168P}. The C$^{18}$O(2-1) emission,
  represented here as one white contour at 0.75 maximum, shows that
  the position angle (PA) of the outer disk is at -20 deg East of
  North, and perpendicular to the solid gray double-arrow, while the
  position angle of the intensity nulls is indicated by the dashed
  double-arrow ($-$8~deg). {\bf b-f:} Radiative transfer prediction
  for polarized intensity in H-band, for different inner disk PAs
  (indicated in degrees on the plots), and for different relative
  inclinations $\alpha$ between the inner and the outer disks, such
  that the net inclination of the inner disk is (--20+$\alpha$)~deg
  relative to the plane of the sky. The $x-$ and $y-$ axis
   indicate offset along RA and DEC, in arcsec.}
\end{figure*}

\section{Results} \label{sec:results}

In order to constrain the PA of the inner disk and $\alpha$, its
inclination with respect to the outer disk, we studied different
orientations while trying to reproduce the shape and position of the
nulls seen in scattered and polarized light. In
Fig.~\ref{fig:warporient} we summarize the radiative transfer
predictions of 5 different configurations. PAs much different from
$-$8 are ruled out, as even a PA of 0 (see
Fig.~\ref{fig:warporient}e) displaces the southern shadow so that it is
inconsistent with the observations. In parallel, for low and negative
$\alpha$, the inclination and orientation of the shadows do not fit
the shape of the nulls (see Fig.~\ref{fig:warporient}c\&f). A
qualitative match with the observations is obtained with an inner disk
inclined at $\alpha = 70$~deg relative to the outer disk, and along a
PA of $-8$deg (Fig.~\ref{fig:warporient}b). We can rule out
configurations with PAs beyond 10~deg of $-8$~deg, so that the
1~$\sigma$ error bar is about 5~deg. Likewise, the relative
inclination is constrained within 60 to 80~deg, so that the 1~$\sigma$
error is also $\sim$5~deg.

An inner disk orientation along a PA of $\sim$60$^{\circ}$ has been
proposed by \citet{Pontoppidan2011ApJ...733...84P}, based on long-slit
spectroscopy of the CO~4.67~$\mu$m line, along with a purely Keplerian
disk model. However, as illustrated in Fig.~\ref{fig:warporient}d,
such an orientation can be discarded from the $H$-band imaging. It is
possible that non-Keplerian kinematics may have biased the orientation
inferred from the ro-vibrational CO.

It is interesting that our models predict a peak $H-$band intensity at
the same position as in the observations, at $\sim$1.5h
(North-North-West).  However, the second peak in the PDI image to the
North-East does not coincide with our radiative transfer
predictions. This can be due to an effect of fine structure in the
outer disk, or to deviations from a perfect ring, or to the stellar
offset from the center of the cavity. These details are beyond the
scope of our model. 

The width of the shadows in the outer disk is dominated by the scale
height of the inner disk, as it covers a wider solid angle of the
star. A more detailed study could lead to a better constraint on the
aspect ratio and flaring of the inner disk.

\section{Discussion} \label{sec:discussion}

Warped disks are found in varied astrophysical contexts. Galactic
warps may be due to a misalignment between a galaxy's angular
momentum and its surrounding dark matter halo
\citep{Binney_1992ARAA..30...51B} or by tidal encounters with nearby
galaxies \citep{Hunter_1969ApJ...155..747H}. \citet{Christiaens2014}
propose that the two-armed spirals in the outer disk of HD~142527
might be indicative of a recent close stellar encounter \citep[see
  also][]{2005AJ....129.2481Q}.  Although a
flyby could also explain the tilt between the inner and outer disks,
no partner for such a stellar encounter has been identified.

In the prototypical T-Tauri disk TW~Hya,
\cite[][]{Rosenfeld2012ApJ...757..129R} proposed a warp to understand
the sinusoidal ($m=1$) azimuthal modulation seen in {\em HST} images,
as well as features of the CO gas kinematics of the inner most
regions. They found that a standard Keplerian disk model was unable to
account for the CO line wings and spatially resolved emission near the
central star, and explored 3 possible interpretations to account for
the observed kinematics: 1) scaling up the temperature by a factor of
three inside the cavity; 2) allowing super-Keplerian tangential
velocities near the star, and 3) invoking a warped disk model in which
the line-of-sight disk inclination increases towards the star. 

%

Dynamical interaction between circumstellar disks and (proto)planets
or sub-stellar companions may lead to warps
\citep[][]{1997MNRAS.292..896M}. In the $\beta$ Pictoris debris disk,
the inner disk warp may have been dynamically induced by $\beta$~Pic~b
\citep[][]{Dawson_2011ApJ...743L..17D}, whose orbit is found to be
aligned with the inclined warped component
\citep[][]{Lagrange_2012AA...542A..40L, Chauvin_2012AA...542A..41C}.

%

A warped inner disk in HD~142527 bears consequences on the physical
conditions in the outer disk. The shadowed regions in the outer disk
behind and along the inner disk midplane can be diagnosed in terms of
temperature decrements in sub-mm continuum imaging
\citep[][]{Casassus2015trap}. There are other interesting consequences
on the dynamics of the system. A question arises as to the origin of
this warped disk, which is probably driven by the low-mass companion
\citep[][]{Biller2012, Rodigas_2014ApJ...791L..37R, Close2014},
although it could also be linked with the disk-envelope interaction
leading to the formation of the binary in the class~I stage. Since
there is no apparent displacement of the shadows by comparing the NICI
images, acquired in June 2011, with the NACO PDI from July 2012, we
can exclude the possibility that the inner tilt is precessing at a
rate faster than $\sim$10~deg~yr$^{-1}$.  Likewise, we can discard
shadows cast by compact concentrations of material in the very closest
vicinity of the star.

The detailed scattered light images of HD~142527 available in the
literature, coupled with state-of-the-art radiative transfer tools,
have allowed us to conclude on the three-dimensional structure of the
system by levering on the inner disk shadows.  The inner regions turn
out to be warped, such that its inclination relative to the outer disk
is 70$\pm$5~deg, and its PA is $-8\pm5$~deg. This finding poses a
challenge to understand the dynamics of the HD~142527 system, and is
an invitation to interpret scattered light images of gapped
protoplanetary disks from the perspective of inner warp shadows.

\acknowledgments

We thank Henning Avenhaus for providing comparison images in FITS
format, and Jorge Cuadra for useful discussions.  SM, SP and SC
acknowledge support from the Millennium Science Initiative (Chilean
Ministry of Economy), through grant ``Nucleus P10-022-F''. SM
acknowledges CONICYT-PCHA/MagísterNacional/2014 - folio 22140628.  SP
and SC acknowledge financial support from FONDECYT grants 3140601 and
1130949, respectively.



\end{document}